\documentclass[twoside]{dis08}
\usepackage[latin1]{inputenc}
\usepackage[dvips]{graphicx,epsfig,color}
\usepackage{wrapfig,rotating}
\usepackage{amssymb,amsmath,array}

\begin{document}
\title{Strange, charm, and bottom flavors\\ 
in CTEQ global analysis\thanks{
Contribution to the Proceedings of XVI International Workshop on
Deep-Inelastic Scattering and Related Subjects (DIS 2008), London, UK,
7-11 April, 2008}
}

\author{Pavel M. Nadolsky$^{1,2}$\\
1- Department of Physics, Southern Methodist University,
Dallas, TX 75275, U.S.A.\\\vspace{.1cm}
2- Department of Physics and Astronomy, Michigan State University,
East Lansing, MI 48824, U.S.A.
}

\maketitle

\begin{abstract}
I discuss advances in the determination of strange, charm, and bottom quark 
parton distribution functions obtained in the CTEQ6.5 and CTEQ6.6 global
analyses. These results affect electroweak precision
observables and certain new physics searches at the Large Hadron
Collider. 
I focus, in particular, on high-energy implications of 
the consistent treatment of heavy-quark threshold effects in DIS 
in the general-mass factorization scheme; an independent parametrization for
the strangeness PDF; and the possible presence of 
nonperturbative (``intrinsic'') charm.
\end{abstract}

\vspace{\baselineskip}
{\bf Introduction.} Treatment of $s$, $c$,
and $b$ quark flavors in the global fit of parton distribution
functions (PDFs) has undergone important developments 
in order to meet demands of modern QCD applications. The recent NuTeV
and CCFR experimental data on charged-current 
deep inelastic scattering directly probe the
strangeness distribution  $s(x)$, allowing it to be independently
determined in the global analysis. 
Dependence of heavy-quark scattering contributions on 
charm- and bottom-quark masses $m_c$ and $m_b$ introduces 
conceptual and practical challenges. Throughout the years, these
challenges were addressed through the development of a general-mass (GM)
factorization scheme \cite{Aivazis:1993pi,Collins:1998rz}, 
an all-order framework for assessment of heavy-quark mass effects 
in the whole kinematical range probed by the PDF analysis. 
The latest CTEQ6.5 \cite{Tung:2006tb,  Lai:2007dq, Pumplin:2007wg} 
and CTEQ6.6 \cite{Nadolsky:2008zw} NLO PDF sets 
provided by our
group are obtained in a new systematic implementation of such
scheme, based on the principles summarized below. The 
new PDFs provide excellent description of the existing data 
in the global analysis, as the previous ones. However, 
the differences due to the improved treatment of mass effects 
give rise to phenomenologically significant shifts in 
certain predictions at the LHC. Implications of these
new developments for collider physics 
are reviewed in two talks at the DIS 2008 workshop  
\cite{NadolskyDIS1,NadolskyDIS2}. 
This contribution summarizes, and further elaborates on, the comments
and figures in the slides for those talks. It is essential to have
Refs.~\cite{NadolskyDIS1,NadolskyDIS2} open while reading this paper.

{\bf Overview of CTEQ6.5 and 6.6 PDFs.} The CTEQ6.5 series of papers
\cite{Tung:2006tb,Lai:2007dq,Pumplin:2007wg}
extended the conventional CTEQ global PDF analysis \cite{Pumplin:2002vw,Stump:2003yu}
to incorporate a comprehensive treatment of heavy-quark effects and
to include the most recent experimental data. The PDFs constructed
in those studies consist of (i) the base set CTEQ6.5M, together with
40 eigenvector sets along 20 orthonormal directions in the parton
parameter space \cite{Tung:2006tb}; (ii)~several PDF sets CTEQ6.5Sn
(n=-2,...4), designed to probe the strangeness degrees of freedom
under the assumption of symmetric or asymmetric strange sea \cite{Lai:2007dq};
and (iii) several sets CTEQ6.5XCn (n=0...6) for a study of the charm
sector of the parton parameter space, in particular, the allowed range
of independent nonperturbative ({}``intrinsic'') charm partons in
several possible models \cite{Pumplin:2007wg}.

The above three publications were followed by the CTEQ6.6 study
\cite{Nadolsky:2008zw}, which incorporated
the free strangeness parametrization $s(x,\mu)$ into the
general-purpose set of 44 error PDFs. (In contrast, the CTEQ6.5 error
  PDFs  assume proportionality of
  $s(x,\mu_0)$ to $\bar u(x,\mu_0) + \bar
  d(x,\mu_0)$ at the initial evolution scale $\mu_0$, 
while free $s(x,\mu)$ and $\bar s(x,\mu)$ were explored in
  separate CTEQ6.5S sets). The CTEQ6.6 set assumes $s(x,\mu_0) = \bar
s(x,\mu_0)$, given that the preference for a non-zero
strangeness sea asymmetry suggested by the NuTeV data remains marginal. 
In addition, we have improved the
numerical computation of heavy-quark contributions to DIS cross
sections, bringing CTEQ6.6M predictions to a better agreement
with DIS heavy-flavor production data ($F_2^c$, $F_2^b$) as
compared to CTEQ6.5M (\cite{NadolskyDIS2}, slide 4). Within this
framework, we provide updated PDFs 
in the ``intrinsic charm'' scenario and for alternative values of
the strong coupling strength, charm and bottom masses
($\alpha_s(M_Z)=0.112-0.125$, $m_c = 1.4$ GeV, $m_b = 4.75$ GeV). 

{\bf Summary of the GM scheme.} Our GM scheme originates in the 
ACOT papers on the factorization for heavy-quark scattering 
\cite{Aivazis:1993pi,Collins:1998rz}. It also includes 
more recent conceptual developments. Its key features
are \cite{Tung:2006tb,ThorneTung:HERALHC}
\begin{itemize}
\item variable number of active quark flavors; 
\item full dependence on the heavy-particle mass ($m_Q$) at energies ($Q$)
  close to the heavy-particle production threshold ($Q\sim m_Q$), for each
  heavy-flavor species;
\item all-order summation of large collinear logarithms $\ln^n(Q/m_Q)$ at
  energies far above the heavy-particle threshold ($Q \gg m_Q$);
\item zero-mass expressions for Feynman graphs
  with initial-state heavy particles (also known as ``flavor-excitation graphs'')
  \cite{Collins:1998rz,Kramer:2000hn}; this feature greatly reduces
  the computational complexity, by evaluating a large fraction of
  heavy-flavor subprocesses with the help of  relatively simple
  zero-mass matrix elements;
\item mass-dependent rescaling of the light-cone momentum fraction in
  flavor-excitation contributions to fully inclusive ($F_{2,3}(x,Q)$)
  and semi-inclusive ($F_{2}^{c,b}(x,Q)$) DIS structure functions 
 \cite{Tung:2001mv}.
\end{itemize}

{\bf Mass thresholds in DIS; quark PDFs at the LHC.} Much of the
latest advancements in the GM framework
focus on kinematical effects in the vicinity of heavy-quark mass
thresholds in inclusive and semi-inclusive DIS. 
As it turns out, these
effects influence both heavy- and light-quark PDFs in a large range
of scattering energies. For example, compare total cross sections
$\sigma_Z$ and $\sigma_W$ for weak
($Z^0$ and $W^{\pm}$)
boson production at the LHC obtained (a) within the GM scheme and  (b)
the common zero-mass (ZM) scheme employed in many 
PDF sets, e.g., in CTEQ6.1 PDFs \cite{Stump:2003yu}. 

The GM CTEQ6.6 $Z$ and $W$ cross sections are larger than the
corresponding ZM CTEQ6.1 cross sections by 6-7\% 
(\cite{NadolskyDIS1}, slide 7; \cite{NadolskyDIS2}, slides
9, 10), which exceeds the magnitude of the NNLO hard-scattering
contribution of order 2\%  \cite{Hamberg:1990np,Harlander:2002wh}, 
as well as the experimentally-driven PDF uncertainty of about 3\%. 
This enhancement reflects the larger magnitude of GM $u$ and $d$
anti-quark PDFs at $x=10^{-3}-10^{-2}$ typical for weak boson
production (\cite{NadolskyDIS2}, slide
8). Despite its modest magnitude, such few-percent 
difference is of import for precision measurements of
$W,Z$ boson cross sections and $W$ boson mass.

To understand the origin of the difference,
notice first that both schemes implement a
variable number $n_f$ of active quark flavors:  they realize 
a sequence of effective factorization schemes with fixed values of 
$n_f$, in which the switching  from the $(n_f-1)$- to $n_f$-flavor 
scheme occurs at a factorization scale $\mu$ of order of the mass 
of the $n_f$'th quark (usually exactly at $\mu = m_{n_f}$). However, 
while the GM scheme retains all
relevant dependence on $m_{c,b}$, the common ZM scheme neglects
this dependence altogether, operating with $n_f$ {\it massless} quarks
when $\mu$ lies between the  $n_f$'th and $(n_f+1)$'th mass
threshold. As a result the common ZM scheme fails to correctly suppress 
the $c$, $b$ contributions to the DIS structure 
functions $F_{\lambda}(x,Q)$ near the respective thresholds, {\it i.e.}, when the DIS total energy 
$W=Q\,(1/x-1)^{1/2}$ is close to $2\,m_c$ or $2\,m_b$. 

In contrast, the GM formalism implements the threshold suppression of 
$F_{\lambda}(x,Q)$ completely (\cite{NadolskyDIS2}, slide 7) 
by including two kinds of contributions dependent on $m_{c,b}$: 
(a) mass-dependent rescaling of the light-cone momentum fraction variable 
in partonic processes with incoming heavy  quarks;
(b) mass-dependent terms in the partonic cross section (Wilson
coefficient) in the light-flavor scattering processes involving 
explicit flavor creation (such as the gluon-photon fusion). 

Since the theoretical
calculations in the global fit must agree with the extensive DIS
data at low and moderate $Q$, the threshold reduction in $c$, $b$, and $g$ 
contributions in the GM NLO fit must be compensated by larger
magnitudes of light ($u$, $d$) quark and antiquark contributions. In the
appropriate $(x,Q)$ region one therefore
sees an increase in the $u$ and $d$ PDFs
extracted in the GM CTEQ6.6 analysis, as compared to those from the ZM CTEQ6.1
analysis.

Although both CTEQ and MRSTW groups have employed some forms of the GM
scheme for many years, the shift in the W and Z cross sections 
brought about by the improved treatment of heavy-flavor effects 
was first noticed in the CTEQ6.5 paper. Subsequent GM global analyses confirm those
findings and converge toward common predictions for $\sigma_{Z,W}$. 
The 2006 \cite{Martin:2007bv} and 2008 \cite{WattDIS} 
MSTW results for $\sigma_{Z,W}$ at the LHC agree 
with CTEQ6.6 within 2-3\%.

{\bf Independent strangeness parametrization.} 
The dimuon DIS data  ($\nu A\to\mu^{+}\mu^{-}X$)
\cite{Tzanov:2005kr} in the CTEQ6.6 fit probe the strange quark distributions via the
underlying process $sW\to c$, making the familiar ansatz
$s(x,\mu_{0})\propto\bar{u}(x,\mu_{0})+\bar{d}(x,\mu_{0})$ unnecessary.
However, as shown in Ref.\,\cite{Lai:2007dq}, the existing
experimental constraints on the strange PDFs remain relatively weak
and have power to determine at most two new degrees of freedom
associated with the strangeness in the limited range $x>10^{-2}$.
At $x\lesssim 10^{-2}$, the available data probe mostly a
combination $(4/9)\left[u(x)+\bar u(x)\right] + (1/9)\left[d(x)+\bar
d(x)+s(x)+\bar s(x)\right]$ accessible in neutral-current DIS, but
not the detailed flavor composition of the quark sea. Therefore, 
the strangeness to non-strangeness ratio at small $x$,
$R_s=\lim_{x\rightarrow
0}\left[s(x,\mu_{0})/\left(\bar{u}(x,\mu_{0})+\bar{d}(x,\mu_{0})\right)\right]$,
is entirely unconstrained by the data, although, on general physics grounds,
one would expect it to be of order 1 (or, arguably, a bit
smaller). Thus, in the current CTEQ6.6 analysis, we adopt a
parametrization for the strange PDF of the form
$s(x,\mu_{0})=A_{0}\, x^{A_{1}}\,(1-x)^{A_{2}}P(x)$, where  
$A_{1}$ is set equal to the analogous parameter of $\bar{u}$
and $\bar{d}$ based on Regge considerations.
A smooth function $P(x)$ (of a fixed form for all 45 CTEQ6.6 PDF sets)
ensures that the ratio $R_s$
stays within a reasonable range (0.63-1.15).

The independence of the strangeness parametrization may affect
predictions for collider observables. For example, 
the ratio $r_{ZW}\equiv\sigma_{Z}/(\sigma_{W^{+}}+\sigma_{W^{-}})$
of the  LHC $Z^{0}$ and $W^{\pm}$ total cross sections is quite sensitive to
the uncertainty in $s(x,\mu)$. Nominally $r_{ZW}$ is an exemplary
``standard candle'' LHC observable, because many common uncertainties 
cancel inside the ratio. This cancellation is
an essential prerequisite for accurate measurements 
of $W$ boson mass \cite{Besson:2008zs}. However, the PDF uncertainty
associated with $s(x,\mu)$  cancels incompletely, in view that it 
contributes to $\sigma_Z$ and $\sigma_W$ through non-identical
subprocesses $s\bar s \rightarrow Z$ and $s c \rightarrow
W$. Since these subprocesses have sizable partial rates 
($\approx$20\% and 27\% at NLO), the correlation between
$\sigma_Z$ and $\sigma_W$  is considerably reduced (and, as a result, the PDF
uncertainty $\Delta r_{ZW}$ on $r_{ZW}$ is increased) if $s(x,\mu)$ is
independent. For instance, $\Delta r_{ZW}$ predicted by CTEQ6.6 PDFs [with independent $s(x,\mu)$] 
is increased threefold as compared to CTEQ6.1 PDFs [with
$s(x,\mu_0)\propto \bar u(x,\mu_0) + \bar d (x,\mu_0)$].
A~plot of
the correlation cosine of $r_{ZW}$ with individual PDFs
(\cite{NadolskyDIS2}, slide 12) confirms that most of $\Delta r_{ZW}$ is 
associated with $s(x,\mu)$ at $0.01<x<0.05$.\footnote{The meaning of the correlation cosine is explained in \cite{Nadolsky:2008zw}.} Hence the independent
parametrization for $s(x,\mu)$,  the least constrained distribution among
the light-quark flavors, is paramount for more realistic estimates of $r_{ZW}$.

{\bf Implications of the ``intrinsic charm''.} While the
general-purpose CTEQ6.6 PDFs generate non-zero charm PDFs  entirely
through perturbative evolution at scales $\mu > \mu_0$, the
``intrinsic charm'' (IC) scenarios implemented in the CTEQ6.6C PDF series  
allow for additional nonperturbative channels for charm production,
leading to $c\!\!\!\!^{^{(-)}}(x,\mu) \neq 0$ at
$\mu=\mu_0$. The IC models implemented in this series are reviewed in  \cite{Pumplin:2007wg}. 

Contrary to the naive perception, IC
is not a purely low-energy phenomenon. An IC-driven enhancement in $c(x,\mu)$
at $\mu \approx m_c$ is preserved by the perturbative 
evolution to the electroweak scale and beyond. The IC may
affect the correlated PDF dependence of the LHC $Z$ and $W$  
cross sections. A  figure showing total cross sections $\sigma_Z$ and
$\sigma_W$ (\cite{NadolskyDIS2}, slide 11) includes predictions 
from two IC models, denoted as ``IC-Sea''
and ``IC-BHPS''. These predictions lie on the verge 
of the CTEQ6.6 error ellipse, indicating 
a potentially non-negligible shift due to IC.  Similar 
IC-driven effects are observed in $Z$, $W$ production at the Tevatron (Fig. 6
in \cite{Nadolsky:2008zw}).  Other charm scattering processes, such as charged
Higgs boson production $c\bar s +
c\bar b \rightarrow H^+$ in 2-Higgs doublet model at the LHC
(\cite{NadolskyDIS2}, slide 15) may be enhanced if IC is
included \cite{Lai:2007dq}. Future measurements
involving charm quarks, such as $pp\!\!\!^{{}^{(-)}} \rightarrow Z c X$, could 
test the mechanism behind charm production, with potential
implications for new physics searches. 

\vspace{\baselineskip}
I thank my coauthors for many discussions of presented results,
Jon Pumplin for helpful comments, 
and Wu-Ki Tung for the critical reading of the manuscript.  
This work and participation in the workshop 
were supported in part by the U.S. National
Science Foundation under awards PHY-0354838, PHY-0555545, the U.S.\ Department of Energy
under grant DE-FG02-04ER41299, Lightner-Sams Foundation,
and by the 2008 LHC Theory Initiative Travel Award. 


\begin{footnotesize}

\end{footnotesize}


\end{document}